\documentclass[aps,prl,twocolumn,groupedaddress,showpacs,amsmath,amssymb,amsfonts]{revtex4}

\usepackage{graphicx}
\usepackage{bm}

\begin{document}
 
 \title{Universal spectral statistics in quantum graphs}
 
 \author{Sven Gnutzmann} \email[]{gnutz@physik.fu-berlin.de}
 \affiliation{Institut f\"ur Theoretische Physik, Freie Universit\"at
   Berlin, Arnimallee 14, 14195 Berlin, Germany}
 
 \author{Alexander Altland} \email[]{alexal@thp.uni-koeln.de}
 \affiliation{Institut f\"ur Theoretische Physik, Universit\"at zu
   K\"oln, Z\"ulpicher Str.\ 77, 50937 K\"oln}
 
 \date{\today}

  \begin{abstract}
    We prove that the spectrum of an individual chaotic quantum graph
    shows universal spectral correlations, as predicted by
    random--matrix theory.  The stability of these correlations with
    regard to non--universal corrections is analyzed in terms of the
    linear operator governing the classical dynamics on the graph.
  \end{abstract}
  
   \maketitle
  
  Fluctuations in spectra of individual complex quantum systems (e.g.
  classically chaotic systems) are universal and can be described by
  the Gaussian ensembles of random--matrix theory (RMT).  This
  statement, promoted to a conjecture by Bohigas, Giannoni and
  Schmit~\cite{BGS}, has been empirically confirmed in numerous
  experimental and numerical analyses~\cite{Guhr,Haake,exception}.
  However, never so far has it been possible to demonstrate
  analytically that spectral fluctuations of {\it
    individual}~\cite{fn_dis} chaotic systems obey RMT statistics.
  Important progress has recently been made within the framework of
  periodic orbit theory: summing over orbit pairs of nearly identical
  action it became possible to prove universality (agreement with the
  predictions of RMT) first for a few~\cite{SR,SR1} and then
  all~\cite{Fritz} coefficients of the {\it short time} expansion of
  the spectral form factor.  Unfortunately it is not known how to
  advance this semiclassical approach into the regime of times larger
  than the Heisenberg time $t_H=\frac{2\pi\hbar}{\Delta E}$ ($\Delta
  E$ is the mean level spacing).
  
  Here, we apply different theoretical concepts to prove universality
  (including the long time regime) for a family of chaotic quantum
  systems, the so--called quantum graphs~\cite{Kottos}.  Quantum
  graphs differ from generic Hamiltonian systems in that they are
  semiclassically exact (The density of states can be represented in
  terms of an exact semiclassical trace formula.), while they do not
  possess an underlying deterministic classical dynamics.  Still they
  display much of the behavior of generic hyperbolic quantum systems;
  equally important, they are not quite as resistant to analytical
  approaches as these.
  
  In previous work, Berkolaiko \textit{et al.}~\cite{greg} developed a
  perturbative diagrammatic language to analyze the semiclassical
  periodic--orbit representation of spectral correlation functions
  beyond the leading ('diagonal') approximation.  In spite of the full
  knowledge of its building blocks~\cite{greg,us} a complete
  resummation of the perturbation series has so far been elusive (not
  to mention that such expansions are subject to the same limitations
  as the semiclassical approaches mentioned above.)  In contrast, our
  present approach avoids diagrammatic resummations.  We rather build
  on two alternative pieces of input, both of which have been
  discussed separately before: \textit{i.)} the exact equivalence of a
  spectral average for a quantum graph with incommensurate bond
  lengths to an average over a certain ensemble of unitary
  matrices~\cite{Kottos,Barra,Tanner}, and \textit{ii.)} the
  so--called color-flavor transformation~\cite{Zirnbauer}, which is an
  (equally exact) mapping of the phase--averaged spectral correlation
  function onto a variant of the supersymmetric $\sigma$-model.  A
  subsequent stationary phase analysis then directly leads to the RMT
  correlation function corresponding to the symmetry of the graph.
  Finally, the spectrum of the 'massive' fluctuations around the
  saddle point contains quantitative information on the stability of
  RMT spectral statistics with regard to non--universal corrections.
  Deferring the discussion of other symmetry classes to a separate
  publication~\cite{ga} we will consider graphs which are invariant
  under both, time reversal and spin rotation.
  
  Let us begin by introducing our basic setting.  A quantum graph
  consists of $V$ vertices $j$ connected by $B$ bonds $b$.  For the
  topology of the graph we assume that pairs of vertices are connected
  by at most one bond and that no bond starts and ends at the same
  vertex \cite{technical}.  We introduce $2B$ double indices $(b,d)$,
  where $d=1,2$ determines the (arbitrarily defined) direction of
  propagation along $b=1,\dots,B$. Boundary conditions on the graph
  are set by the fixed $2B$--dimensional unitary matrix $S_{bd,b'd'}$
  which describes the scattering of an incoming wave function on bond
  $b$ to an outgoing wave function on bond $b'$.  Of course,
  $S_{bd,b'd'}$ is non--vanishing only for bonds $b$ and $b'$
  connecting at a common vertex $j$.  Time--reversal invariance
  (${\cal T}$--invariance) implies that $S^T = \sigma_1^{\rm dir} S
  \sigma_1^{\rm dir}$, where $\sigma_i^{\rm dir}=(\sigma_i^{\rm
    dir})_{dd'}$ are Pauli matrices in the space of directional
  indices.  The complete dynamical information on the graph is carried
  by the $2B \times 2B$ bond scattering matrix $\mathcal{S}(k)=T(k) S
  T(k)$.  Here, the diagonal matrices $T(k)$ contain the dynamical quantum
  phases picked up during propagation at fixed wave number $k$ along
  the bonds: $T(k)_{bb',dd'}= \delta_{bb'}\delta_{dd'} 
  \exp{i\frac{k L_b}{2}}$, where $L_b$
  is the length of bond $b$ and the two--fold replication in direction
  space expresses the independence of the dynamical phases on the
  direction of propagation.  The concise formulation of this fact
  reads as ${\cal T}:\; \mathcal{S}(k)= \sigma_1^{\mathrm{dir}}
  \mathcal{S}^T(k)\sigma_1^{\mathrm{dir}}$.  The prime signature of
  chaotic dynamics on the graph are strong non--Poissonian
  correlations in its discrete spectrum $\{k_n\}$. The latter is
  defined by the condition that $\mathcal{S}(k_n)$ has a unit
  eigenvalue or, equivalently, by the vanishing of $\xi(k)\equiv
  \det(1-{\cal S}(k))$ at $k=k_n$.  (This condition is
  equivalent\cite{Kottos} to the existence of an eigenvalue of the
  bond Schr\"odinger operator canonically associated to the scattering
  matrix. In this sense the spectrum $\{k_n\}$ is analogous to the
  discrete energy spectrum of a Hamiltonian chaotic system.)  Below,
  we will explore the two--point spectral correlation function
  $$R_2(s)\equiv \Delta^2 \langle \rho(k+s\Delta)\rho(k)\rangle_k-1$$
  where $\rho(k)\equiv\sum_n \delta(k-k_n)$ is the spectral density
  and $\langle \dots \rangle_k\equiv\lim_{K\to \infty}{1\over
    K}\int_0^Kdk(\dots)$ an average over the wave number parameter
  $k$. ($\Delta=\pi/B\bar{L}$ the average level spacing and
  $\overline{L}=\sum_b L_b/B$ the mean bond length.)
  
  Universal behavior of the spectral correlation function (agreement
  with the prediction $R_2^{\rm RMT}$ of RMT) can be expected if the
  corresponding classical system is chaotic (hyperbolic). What is the
  equivalent condition on a quantum graph?  An answer has been
  formulated by Tanner~\cite{Tanner} (see also~\cite{gregongap}) in
  terms of the classical probability $F_{bd, b'd'}\equiv |S_{bd,
    b'd'}|^2= |\mathcal{S}(k)_{bd, b'd'}|^2 $ to get from $(b',d')$ to
  $(b,d)$. This 'classical propagator' $F$ has one eigenvalue
  $\lambda_1=1$, corresponding to equidistribution in bond space. The
  dynamics is \emph{mixing} if, for large times, any initial
  probability distribution converges to this distribution, i.e.\ 
  $\lim_{n\rightarrow\infty} (F^n)_{bd, b'd'}=\frac{1}{2B}$.  This
  condition is met if all other eigenvalues $|\lambda_{2,\dots,B}|<1$
  lie inside the complex unit circle. However, mixing dynamics alone
  does not suffice to guarantee universality of a quantum
  graph\cite{gregonstars}.  An additional condition proposed by
  Tanner\cite{Tanner} states that in the limit $B\to \infty$, the
  spectral gap $\Delta_\mathrm{g}={\rm
    max}_{b\in\{2,\dots,B\}}(1-|\lambda_b|)$ be constant or, at least,
  vanish slowly enough $\Delta_g\sim B^{-\alpha}$. Building on the
  so--called diagonal approximation (an approximation that obtains its
  asymptotics of $R_2(p)$ for large values of $p$) Tanner conjectured
  that universal behavior should be expected for values of the gap
  exponent $0\le \alpha <1$.
  
  In the following we will show that quantum graphs indeed show RMT
  spectral correlations (provided the condition $0\le \alpha< 1/2$
  somewhat stronger than Tanner's is met.)  We start out from the
  representation $\rho(k)=\Delta^{-1}-{1\over \pi}\frac{d}{dk}{\rm
    Im}\ln \xi(k^+), k^+ \equiv k+ i0$ of the density of states in
  terms of the spectral determinant\cite{Kottos}. Using this formula,
  it is straightforward to verify that the two--point function assumes
  the form $R_2(p)={1 \over 8\pi^2}{d^2\over dj_+ dj_-}\big|_{j=0}{\rm
    Re\,} \langle \zeta(j_+,j_-)\rangle_k$, where
  \begin{equation}
    \label{eq:1}
    \zeta(j_+,j_-) \equiv{\xi(k^+ +p_{+\mathrm{\bf f}})
      \over \xi(k^+ +p_{+\mathrm{\bf b}})}
    \left(\xi(k^+ +p_{-\mathrm{\bf f}})\over 
      \xi(k^+ +p_{-\mathrm{\bf b}})\right)^*
  \end{equation}
  and $p_{\pm\mathrm{\bf b}}=(\pm s/2 - j_\pm)\Delta$,
  $p_{\pm\mathrm{\bf f}}=(\pm s/2 + j_\pm)\Delta$ .
  
  Our analysis will be based on the assumption that all bond lengths
  $L_b$ are rationally independent.  It has been shown~\cite{Barra}
  that under this condition the average over the parameter $k$ is
  strictly equivalent to an average over $B$ independent phases
  $e^{i\frac{kL_{b}}{2}}\mapsto e^{i \phi_{b}}$:
  \begin{equation}
  \label{eq:equiv}
    \langle \mathcal{F}[T(k)]\rangle_k
    =
    \langle \mathcal{F}[T(\phi)] \rangle_{\phi},
  \end{equation}
  where ${\cal F}$ is a smooth function of the bond--diagonal phase
  matrix $T$ introduced above.
  
  To prepare the application of (\ref{eq:equiv}) to our present
  problem, we represent the fraction of determinants in (\ref{eq:1})
  as a Gaussian integral, $\zeta = \int d(\bar\psi,\psi) \langle
  \exp(-{\bm{S}}[\bar\psi,\psi])\rangle_\phi$, where~\cite{sdet}
  \begin{equation}
    \label{eq:retarded_action}
    \bm{S}[\bar{\psi} \!,\psi]\!=\! \bar\psi_+\! \!\!
    \left[
      \begin{smallmatrix}
        1 & \!\!\! T(k)\\
        T(k) &(ST_+)^\dagger
      \end{smallmatrix}\right]
    \!\! \psi_+ \!+\! \bar\psi_-
    \!\!\!
    \left[
      \begin{smallmatrix}
        1 & \!\!\!T(k)^\dagger\\
        T(k)^\dagger &S T_-
      \end{smallmatrix}\right]
    \!\!\psi_-.
  \end{equation}
  Here, $\psi=\{\psi_{a,s,x,d,b}\}$ is a $16B$-dimensional
  supervector, where $a=\pm$ distinguishes between the retarded and
  the advanced sector of the theory (determinants involving ${\cal S}$
  and ${\cal S}^\dagger$, resp.), $s=\mathrm{\bf f},\mathrm{\bf b}$
  refers to complex commuting and anti--commuting components
  (determinants in the denominator and numerator, resp.), and $x=1,2$
  to the internal structure of the matrix kernel appearing in
  (\ref{eq:retarded_action}).  Defining $\sigma_i^{\rm bf}$ as the
  Pauli--matrices in superspace, the matrices $T_\pm \equiv T(2
  p_\pm)$ i.e. $T_\pm$ are diagonal matrices in superspace containing
  the bond matrices $T\left(2p_{\pm,\mathrm{\bf b/f}}\right)$ in the
  boson--boson/fermion-fermion sector.  Using that
  $\xi(k+p)=\det(1-{\cal S}T(2p))$ and $\det(1- {\cal S}T(2p)) =
  \det\left(\begin{smallmatrix} 1& T(k) \cr T(k) & (ST(2p))^\dagger
  \end{smallmatrix}\right)$, one verifies that the Gaussian
integration over all components of $\psi$ indeed yields the
determinant $\zeta$.
  
As a second step, we subject the phase--averaged $\psi$--functional to
a duality transformation known as the color--flavor
transformation~\cite{Zirnbauer}. In a variant adapted to the present
context~\cite{technicalnote}, the transformation states that
  \begin{equation}
    \label{eq:CFtrafo}
     \langle \exp(-\bm{S}[\bar\psi,\psi])\rangle_\phi=
  \langle \exp(-\bm{S'}[\bar\Psi,\Psi])\rangle_Z,
  \end{equation}
  where $\langle \cdot \rangle \equiv \int dZ d \tilde Z{\,\rm
    sdet\,}(1-Z\tilde Z) (\cdot)$ and (matrix structure in
  advanced/retarded space)
  \begin{equation}
    \label{eq:SpsiS}
    \bm{S'}[\bar\Psi,\Psi]=\bar\Psi_1 \left[
      \begin{smallmatrix}
        1& Z\vphantom{(ST_+)^\dagger} \cr
        Z^{\tau}\vphantom{(ST_+)^\dagger} & 1
      \end{smallmatrix}\right]\Psi_1 + \bar \Psi_2\left[
      \begin{smallmatrix}
        (ST_+)^\dagger & \tilde Z^{\tau} \cr \tilde Z & ST_-
      \end{smallmatrix}\right]\Psi_2.
  \end{equation}
  Referring~\cite{technicalnote} for a short discussion of the
  underlying technicalities, we here briefly explain the notation and
  the physical meaning of the transformation (\ref{eq:CFtrafo}).  In
  (\ref{eq:SpsiS}), $\Psi_{1,2}=\{ (\Psi_{1,2})_{a,s,t,b,d}\}$ are
  $16B$--dimensional independent supervectors, where the index,
  $t=1,2$ accounts for the time--reversal symmetry of the model.
  Presently, all we need to know about the variables $\Psi$ and $\bar
  \Psi$ is that they contain elements of $\psi$ and $\bar \psi$ as
  their components, and fulfill $\bar \Psi_{1,2}=\Psi_{1,2}^T \tau$.
  Here, we introduced the fixed supermatrix $\tau\equiv
  \sigma_1^{\mathrm{dir}} \otimes \tau_0$, where $\tau_0 \equiv
  \left(E_{\mathrm{\bf bb}} \sigma_1^{\mathrm{ tr}}-i E_{\mathrm{\bf
        ff}} \sigma_2^{\mathrm{tr}}\right)$ ($\sigma_i^{{\rm tr}/{\rm
      dir}}$ are Pauli matrices in the 'time--reversal' index $t$ and
  direction index $d$, $E_{\mathrm{\bf bb/ff}}$ are projectors on the
  bosonic/fermionic sector of the theory). The newly introduced
  integration variables, $Z={\,\rm bdiag\,}(Z_1,\dots,Z_B)$ are
  $8B$--dimensional block--supermatrices with $8$--dimensional entries
  $Z_b=\{Z_{b,ss',dd',tt'}\}$. Finally, $Z^{\tau} \equiv \tau Z^T
  \tau^{-1}$ is, in a generalized way, transposed to $Z$, while $Z$
  and $\tilde Z$ are independent.

  \begin{figure}[hbt]
    \centerline{\resizebox{6cm}{!}{\includegraphics{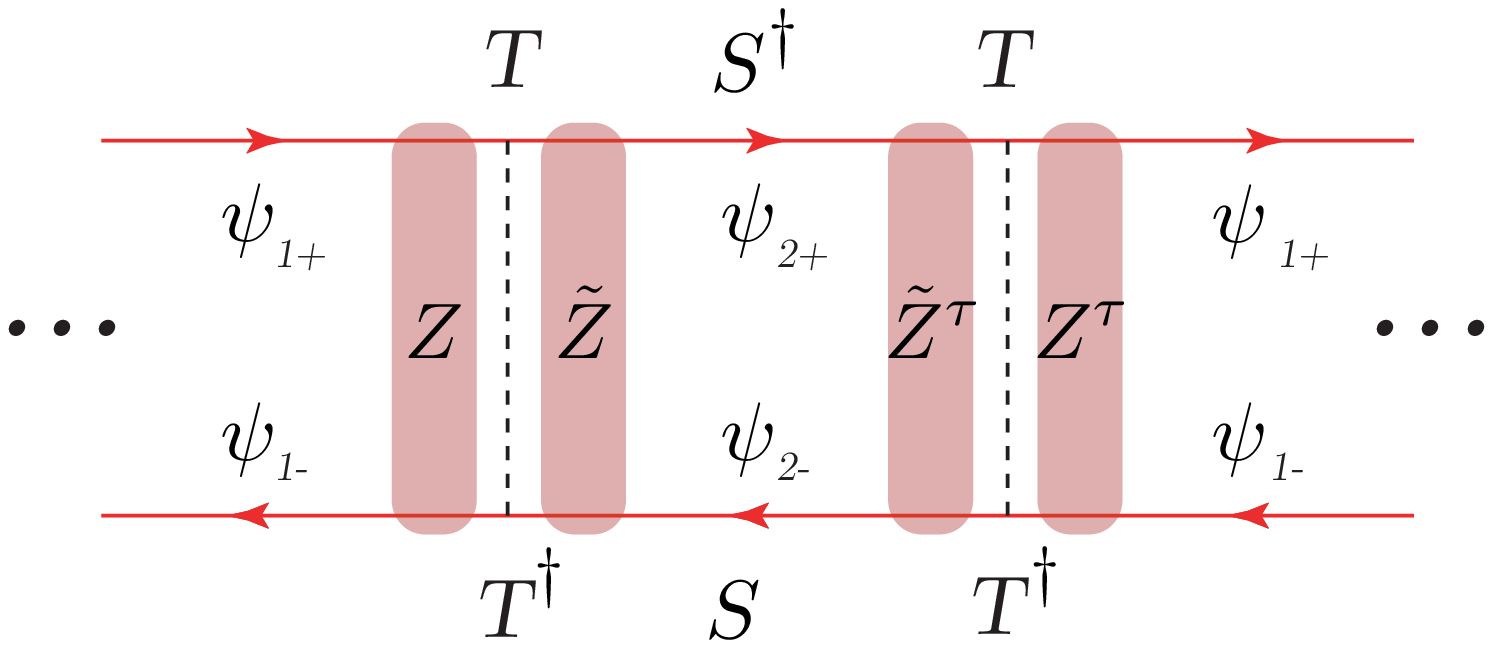}}}
    \caption{On the physical interpretation of the color-flavor
      transformation. Explanation, see text.}
    \label{Zmode}
  \end{figure}

  What is the physical significance of the transformation
  (\ref{eq:CFtrafo})? Fig. \ref{Zmode} shows a cartoon of the retarded
  (upper line) and advanced (lower line) wave function dynamics in the
  system.  During propagation, both states pick up random scattering
  phases $T$ (indicated by vertical dashed lines) and suffer
  scattering from one bond to the other ($S$--matrix). The rapid
  succession of these events implies wild fluctuations of the wave
  function amplitudes. Within the field theoretical context, this
  translates to uncontrollable fluctuations of the bilinears
  $\bar\psi_{+,s,x,d,b}e^{i\phi_b}\psi_{+,s,x,d,b}$ appearing in the
  action (\ref{eq:retarded_action}). In contrast, the field $Z$ enters
  the theory as $\sim \bar
  \Psi_{+,s,t,d,b}Z_{b,ss',tt',dd'}\Psi_{-,s',t',d',b}$, i.e. through
  structures that {\it couple} retarded and advanced field amplitudes
  (the 'vertical' ovals in the figure). These amplitudes generally
  {\it interfere} to form slowly fluctuating entities (the basic
  principle behind the formation of universal correlations.)  This
  indicates that the $Z$--integral will be comparatively benign and,
  foreseeably, amenable to stationary phase approximation schemes.  To
  promote this expectation to a quantitative level, we integrate out
  the $\Psi$'s, thus arriving at the \emph{exact} representation
  $\langle \zeta \rangle_k = \int dZ d \tilde Z \exp(-\bm{S}[Z,\tilde
  Z])$,
  \begin{equation}
    \begin{split}
      \bm{S}[Z,\tilde Z]=& - \mathrm{str\; ln}
      \left(1-\tilde{Z}Z\right)
      +\frac{1}{2} \mathrm{str\; ln} \left(1-Z^{\tau} Z\right)\\
      & +\frac{1}{2} \mathrm{str\; ln} \left(1-(ST_-)^\dagger
        \tilde{Z} (ST_+) \tilde{Z}^{\tau} \right).
    \end{split}
    \label{eq:fullaction}
  \end{equation}
  As a first step towards a better understanding of the physics of
  this expression, let us consider its quadratic expansion,
  \begin{equation}
  \label{eq:S2}
    \bm{S}^{(2)}[Z,\tilde Z]\!=\!\mathrm{str}\left(\tilde{Z}Z-\frac{1}{2}
  Z^{\tau} Z -
    \frac{1}{2}T_-^\dagger S^\dagger  \tilde{Z} ST_+
  \tilde{Z}^{\tau}\right).
  \end{equation}
  Void of non--linearities (terms of ${\cal O}(Z^4)$), the action
  $\bm{S}^{(2)}$ describes the un--interrupted propagation of two
  amplitudes along the {\it same} path in configuration space, i.e.
  the level of approximation underlying the diagonal approximation in
  semiclassical periodic--orbit theory.
  
  This connection is made quantitative by noting that the action
  $\bm{S}^{(2)}$ possesses a family of approximately (up to
  corrections of ${\cal O}(B^{-1})$) 'massless' configurations, or
  'zero modes' identified by $\delta_Z \bm{S}^{(2)}=\delta_{\tilde Z}
  \bm{S}^{(2)}=0$. Upon substitution of the ansatz
  $Z_{dd'}=\delta_{dd'} Z_d$ (configurations not diagonal in direction
  space do not qualify as solutions, see the discussion of deviations
  below) these equations assume the form
  \begin{equation}
    \label{eq:MFsol}
    \tilde Z = Z^{\tau},\qquad (\openone - \hat F)Z=0,
  \end{equation}
  where we have set the parameter matrices $T_\pm=\openone$\cite{fn2}.
  Owing to the fact that on a chaotic quantum graph the 'classical
  propagator' $\hat F$ has only one eigenvalue 1, Eq.
  (\ref{eq:MFsol}) possesses the unique solution
  $Z_{b,dd'}=(2B)^{-1/2}\delta_{d,d'} Y$, proportional to the
  invariant equidistribution. We define $Y^{\tau} = \tau_0 Y^T
  \tau_0^{-1}$, where the matrix $\tau_0$ differs from $\tau$ by the
  absence of the (now redundant) matrix $\sigma_1^{\rm dir}$.
  Technically, the relation $\tilde Y = Y^\tau$ identifies $(Y,\tilde
  Y)$ as generators of the orthosymplectic algebra ${\rm osp}(4|4)$.
  
  To explore the contribution of these modes to the theory, we set
  $T_{\pm,b}= \openone+ i L_b p_\pm$ (provided the bond length
  fluctuations are not too large, $L_b/\overline{L} ={\cal O}(B^{0})$,
  higher orders in the expansion may safely be neglected.), do the
  Gaussian integral over $Y$ and differentiate w.r.t. the source
  parameters $j_{\pm}$.  As a result we obtain
  $R_2^\mathrm{GOE,diag}=1/(\pi^2 s^2)$ which agrees with the $s\gg 1$
  asymptotics of the RMT correlation function.  This is consistent
  with the observation that non--Gaussian contributions to the
  expansion of $\bm{S}[Y]$ can no longer be neglected once $s\lesssim
  1$.
  
  However, before going beyond the level of the Gaussian
  approximation, let us briefly consider the role of non--zero mode
  fluctuations.  A glance at Eq.  (\ref{eq:S2}) shows that deviations
  from the first of the two equations in (\ref{eq:MFsol}) are
  penalized by a large action $\bm{S}^{(2)}={\cal O}(1)$. Upon
  integration, these modes produce a factor unity to the spectral
  determinant. Similarly, due to the absence of multiple connectivities
  and vertex loops, modes that are off--diagonal in the
  direction index $d$ can be integrated out to give a factor unity.
  To explore the more interesting role played by deviations from the
  equation $(\openone-\hat F)Z=0$, let us expand a general
  configuration $Z_{b, dd'} = \delta_{dd'}
  \sum_{m=1}^{2B} Y_m \chi_{m,bd}$ in the basis of eigenfunctions
  $\chi_m$ of the operator $\hat F$.  Here, $Y_m$ are
  four--dimensional supermatrices obeying the symmetry $\tilde Y_m =
  Y_m^\tau$ and the identification $Y_1 \equiv Y$ is understood.  For
  $T_\pm=\openone$ the action of the $2B-1\sim B$ modes $Y_{m>1}$ is
  given by $ \frac{1}{2}\sum_{m} (1-\lambda_m)
  \mathrm{str}\tilde{Y}_mY_m\ge{\Delta_{g}\over
    2}\sum_m\mathrm{str}\tilde{Y}_mY_m$.  Gaussian integration over
  these modes obtains a contribution $\sim \sum_m
  (\Delta/(1-\lambda_m))^{2}\sim 1/\Delta_g^2 B$ to the spectral
  function.  We conclude that the cumulative contribution of the
  `massive' modes can safely be neglected provided $B\Delta_g^{2}\to
  \stackrel{B\to\infty}{\rightarrow} 0$ \cite{fn_massive}. (The same
  holds true for higher order correlation functions, provided the
  order of the function is smaller than $B$.)
  
  Going beyond the level of the quadratic approximation, we note that
  the saddle point equations $\delta_Z \bm{S} = \delta_{\tilde
    Z}\bm{S} =0$ of the full action (\ref{eq:fullaction}) are still
  solved by the zero mode configurations (\ref{eq:MFsol}).  While
  deviations from the zero modes continue to be negligible (as long as
  $B\Delta_g^2 \to \infty$), the action of the latter now reads
  $$\bm{S}[Y]= i B \overline{L}\, \mathrm{str} \left( {p_- \tilde{Y}Y
      /( 1-\tilde{Y} Y)}- {p_+ Y\tilde{Y}/( 1-Y\tilde{Y})}\right),$$
  where we have rescaled $Y\to (2B)^{1/2} Y$.  To represent this
  result in a perhaps more widely recognizable form, let us define the
  $8\times 8$--matrix $ Q=\left(\begin{smallmatrix} 1&Y\\ \tilde Y & 1
  \end{smallmatrix}\right) \left(\begin{smallmatrix} 1&0 \\ 0 & -1
  \end{smallmatrix}\right) \left(\begin{smallmatrix} 1&Y\\ \tilde Y &
    1 \end{smallmatrix}\right)^{-1}.  $ It is then straightforward to
verify that the action $S[Y]$ assumes the form of
Efetov's\cite{Efetov} action for the GOE correlation function
  \begin{equation}
    \label{eq:SEfetov}
    \bm{S}[Q] = {\frac{i \pi}{2}} {\,\rm str\,}(Q\hat \epsilon),
  \end{equation}
  where $\hat \epsilon=-{\rm diag}(p^+_\mathrm{\bf b}, p^+_\mathrm{\bf
    f},p^-_\mathrm{\bf b}, p^-_\mathrm{\bf f})/\Delta$ and the
  integration measure $dQ=d\tilde{Y}dY$.
  
  Summarizing, we have proven Tanner's conjecture on universal
  spectral statistics in large chaotic quantum graphs. Our analysis
  was based on the assumption of (i) incommensurate bond lengths, (ii)
  the absence of multiple connectivities (iii) moderate bond length
  fluctuations, $L_b/\bar L<\cal{O}(B)$, and (iv) weak scaling of the
  spectral gap $\Delta_{g} \sim B^{\alpha}, 0\le \alpha<1/2$ of the
  `classical propagator' on the graph (which is
  stronger than Tanner's expectation $0 \le \alpha<1$.) The conditions
  (i)-(iv) are met by various families of graphs \cite{greg, Tanner,
    gregongap} including the class of fully connected graphs with
  Neumann boundary conditions.

  \begin{acknowledgments}
    We have enjoyed fruitful discussions with Fritz Haake, Sebastian
    M\"uller, Stefan Heusler, and Peter Braun. This work has been
    supported by SFB/TR12 of the Deutsche Forschungsgemeinschaft.
  \end{acknowledgments}


\begin{thebibliography}{}
\bibitem{technical} These technical assumptions simplify the
    analysis. We believe them to be physically irrelevant.
\bibitem{fn2} This simplification is justified because in practice
    (see below) $T_\pm=\openone+{\cal O}(B^{-1})$.
\bibitem{fn_massive} In deriving this result, we have neglected the
    coupling of the modes introduced by $T_\pm$.  However, for
    $L_b/\overline{L}=\mathcal{O}(B^0)$, this coupling leads to
    corrections to the action of ${\cal O}(B^{-1})$ which are
    negligible.
\end{thebibliography}
\end{document}